\documentclass[10pt,superscriptaddress,twocolumn,amsmath,amssymb,aps,prl]{revtex4}
\usepackage{mathrsfs}
\usepackage{graphicx}
\usepackage{dcolumn}
\usepackage{bm}

\newcommand{\be}{\begin{equation}}
\newcommand{\ee}{\end{equation}}
\newcommand{\bea}{\begin{eqnarray}}
\newcommand{\eea}{\end{eqnarray}}

\begin{document}

\title{ A Possible Family of  Ni-based High Temperature Superconductors}

\author{Congcong Le}
\affiliation{Beijing National Laboratory for Condensed Matter Physics,
and Institute of Physics, Chinese Academy of Sciences, Beijing 100190, China}
\affiliation{Kavli Institute of Theoretical Sciences, University of Chinese Academy of Sciences,
Beijing, 100190, China}

\author{Jinfeng Zeng}
\affiliation{Beijing National Laboratory for Condensed Matter Physics,
and Institute of Physics, Chinese Academy of Sciences, Beijing 100190, China}
\affiliation{University of Chinese Academy of Sciences, Beijing 100049, China}

\author{Guang-Han Cao}
\affiliation{Department of Physics, Zhejiang University, Hangzhou 310058, China}

\author{Jiangping Hu}\email{jphu@iphy.ac.cn}
\affiliation{Beijing National Laboratory for Condensed Matter Physics,
and Institute of Physics, Chinese Academy of Sciences, Beijing 100190, China}
\affiliation{Kavli Institute of Theoretical Sciences, University of Chinese Academy of Sciences,
Beijing, 100190, China}
\affiliation{Collaborative Innovation Center of Quantum Matter,
Beijing, China}

\date{\today}

\begin{abstract}
We  suggest that  a family of Ni-based compounds, which contain  [Ni$_2$M$_2$O]$^{2-}$(M=chalcogen)  layers with an antiperovskite structure constructed by mixed-anion Ni complexes, NiM$_4$O$_2$,  can be  potential high temperature superconductors upon doping or applying pressure.  The layer structures  have been formed in many  other transitional metal compounds such as La$_2$B$_2$Se$_2$O$_3$(B=Mn, Fe,Co). For the Ni-based compounds, we predict that the parental compounds  host   collinear antiferromagnetic states similar to those in the iron-based  high temperature superconductors.  The electronic physics near Fermi energy is controlled by two e$_{g}$ d-orbitals with completely independent in-plane kinematics.   We predict that the superconductivity in this family is characterized by strong competition between extended s-wave and d-wave pairing symmetries.
\end{abstract}

\pacs{74.20.Mn, 74.70.Dd, 74.20.Rp}

\maketitle
Since the discovery of cuprates\cite{Cu}, the Cu-based high temperature superconductors, more than thirty years ago,  there have been intensive efforts to find Ni-based counterparts\cite{Anisimov1999,Pickett2004,Lacorre1992,Poltavets2009} as Ni is the nearest neighbor element to Cu  among  the 3d transition metal elements  in the Period Table.  However, although numerous  discovered Ni-based compounds share  similar physics in a variety of aspects to cuprates, none of the known  Ni-based materials exhibits high T$_c$ superconductivity.

  Recently, we have suggested that there is a direct roadmap to design possible high T$_c$ materials\cite{jpHuprx, jpHusciencebult}. In order  to achieve  unconventional high T$_c$, it is necessary to  have an  electronic structure in which those d-orbitals of  transition metal atoms with the strongest in-plane coupling to the p-orbitals of anions have to be isolated near Fermi energy.  In such an electronic structure, the superexchange antiferromagnetic interactions can be maximized to provide superconducting pairing.  Both cuprates and the recently discovered iron-based superconductors\cite{Fe}  are shown to satisfy this condition. Specifically,  in the perovskite-type of structure such as cuprates, the $d_{x^2-y^2}$ e$_g$ orbital  can only be isolated near the d$^{9}$ filling configuration at Cu$^{2+}$,  and in the iron-based superconductors,    the d$^6$ configuration of Fe$^{2+}$ is an unique configuration to  isolate the t$_{2g}$ orbitals near Fermi energy\cite{jpHuprx, jpHusciencebult}.  More importantly, we have pointed out that  such an electronic environment exists  very rarely in nature  because of symmetry and chemistry reasons. Thus, the condition can be considered as the gene of   unconventional high T$_c$ superconductors  to serve as a guide to search for or design high T$_c$ materials.    Following this understanding, we have predicted that there are two specific cases in which the condition can be satisfied with a d$^7$ filling configuration, namely,   Co$^{2+}$-based compounds\cite{jpHuprx,Hu-Co2017,Le-BaCoSO-2017}.

The d-orbital filling configuration  in Ni atoms is d$^8$.  In the d$^8$ configuration,  it is difficult to design a structure to meet the above condition. The reasons are as follows. With the  even filling configuration,  similar to iron-based superconductors, it is necessary to isolate two near-degenerated orbitals  at  Fermi energy and both of them  should   strongly couple to in-plane  p-orbitals. The isolation requires a large energy separation between the selected two orbitals and the rest.  The octahedra complex is the only complex structure to achieve large energy separation in which  the two e$_g$ orbitals have much higher energy than the three t$_{2g}$ orbitals.  Unfortunately, in the  conventional perovskite-type structure, the two e$_{g}$ orbitals have completely different in-plane kinematics as the $d_{z^2}$ orbital has little in-plane coupling to p-orbitals.  These facts can explain why it is difficult for Ni-based materials to achieve high T$_c$ superconductivity.

\begin{figure}[tb]
\centerline{\includegraphics[width=0.45\textwidth]{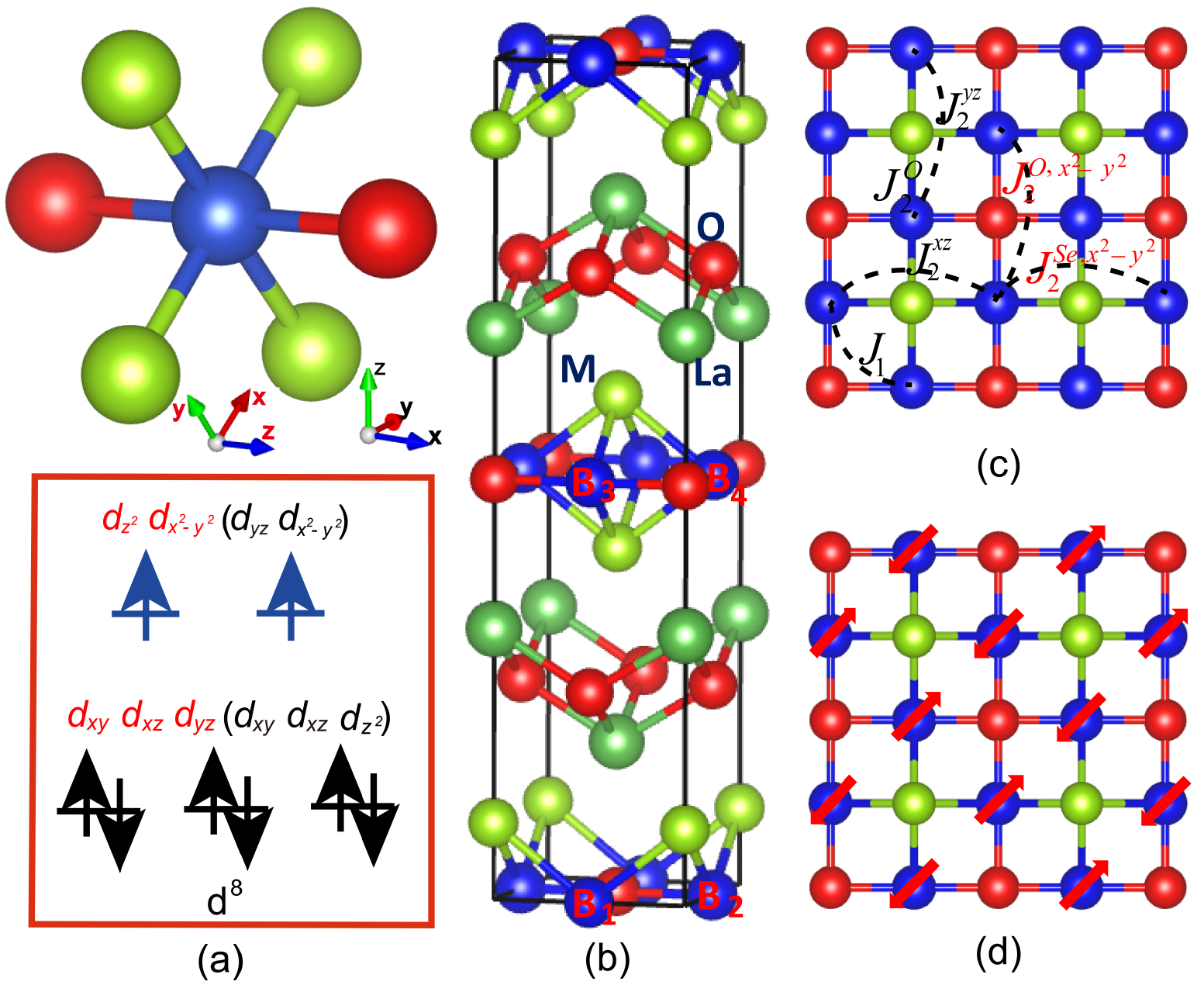}}
\caption{(color online). (a) The illustration of the BM$_4$O$_2$ complex under two different axis coordinations and the energy of the crystal field together with orbital characters; (b) The crystal structure of La$_2$B$_2$Se$_2$O$_3$; (c) The B$_2$M$_2$O layer in the $ab$ plane in which magnetic exchange interactions between NN $J_1$, NNN $J^{xz}_2$, NNN $J^{yz}_2$, NNN $J^{O,x^2-y^2}_2$ and NNN $J^{Se,x^2-y^2}_2$ are indicated; (d) show the C-type collinear AFM state.}
\label{fig1}
\end{figure}

In this letter, we show that it is possible to make both e$_{g}$ orbitals to strongly participate in-plane kinematics in a structure with mixed anion Ni-octahedra complexes, NiM$_4$O$_2$ as shown in Fig.\ref{fig1}(a). The idea is to rotate the complex  and connect them so that the apical oxygens can form a square lattice as shown in Fig.\ref{fig1}(c). In this case, the layered sheets of [B$_2$M$_2$O]$^{2-}$ compose of face-sharing tilted Ni$_2$M$_2$O octahedra where the Ni  atom is surrounded by two axial oxygen atoms and four M atoms. The two $d_{z^2}$ and $d_{x^2-y^2}$ e$_g$ orbitals before the rotation are labeled as $d_{x^2-y^2}$ and $d_{xz/yz}$ orbitals in the new axis coordination as shown in Fig.\ref{fig1}(a). The new  $d_{x^2-y^2}$    gains in-plane kinematics through oxygens  and the new $d_{xz/yz}$  strongly couples to  M anions and maintains in-plane kinematics through M anions.  The  in-plane kinematics of the two orbitals are  completely decoupled because of the in-plane mirror symmetry.


We demonstrate the above idea in  the layered  compounds, La$_2$Ni$_2$Se$_2$O$_3$  as shown in Fig.\ref{fig1}(b),  which are composed of the [Ni$_2$Se$_2$O]$^{2-}$ layers. It is found that the superexchange antiferromagnetic(AFM) exchange couplings are maximized in the Ni-based compound to form a collinear AFM state, the same magnetic state in the parental compounds of   iron-based superconductors\cite{DaiPengchengRMP2015}. The low energy electronic physics is controlled entirely  by the two e$_g$ orbitals which form two independent electronic band structures. Considering that  the superconducting pairing originates from the AFM superexchange interactions, we predict that the compound is characterized by the strong competition between d-wave and extended $s$-wave pairing symmetries.  While the extended s-wave is favored upon hole doping, the $d$-wave can become highly competitive under electron doping or by adjusting lattice parameters, which can lead to a rich phase diagram to include possible time reversal symmetry breaking pairing states.

We first use density functional theory(DFT) to investigate the magnetism and electronic structures  of La$_2$B$_2$Se$_2$O$_3$ (B=Fe,Co,Ni)  which has the space group $I4/mmm$. This type of calculation has been successfully applied to other electron-electron correlated systems.  In particular, the calculation can qualitatively predict both the electronic structures and magnetic orders in the different family of iron-based superconductors\cite{Kurohi2008,AnJiming2009,Zeng2017}.
Our calculations are performed using density functional theory (DFT) as implemented in the Vienna ab initio simulation package (VASP) code \cite{Kresse1993,Kresse1996,Kresse1996B}. The Perdew-Burke-Ernzerhof (PBE) exchange-correlation functional and the projector-augmented-wave (PAW) approach are used. Throughout the work, the cutoff energy is set to be 550 eV for expanding the wave functions into plane-wave basis. In the calculation, the BZ is sampled in the k space within Monkhorst-Pack scheme\cite{MonkhorstPack}. On the basis of the equilibrium structure, the k mesh used is $10\times10\times3$. We relax the lattice constants and internal atomic positions with GGA, where the plane wave cutoff energy is 600 eV. Forces are minimized to less than 0.01 eV/\AA in the relaxation. The GGA plus on-site repulsion $U$ method (GGA$+U$) in the formulation of Dudarev {\it et al}.\cite{Dudarev1998} is employed to describe the electron correlation effect.

\begin{figure}[tb]
\centerline{\includegraphics[width=0.4\textwidth]{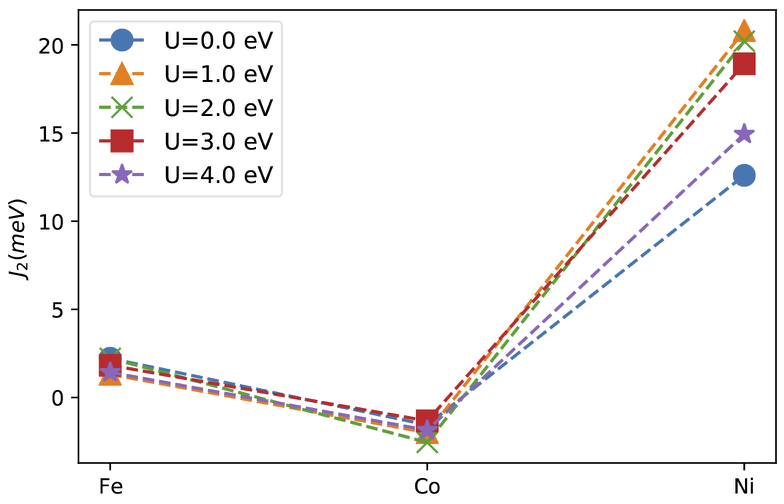}}
\caption{(color online). The average superexchange AFM interaction strength (namely, the NNN AFM, J$_2$)  in  La$_2$B$_2$Se$_2$O$_3$ (B=Fe, Co, Ni), which are extracted from the GGA + U
calculations with the values U = (0.0, 1.0, 2.0, 3.0, 4.0) eV.}
\label{AFM}
\end{figure}

The Mn, Co and Fe compounds have been experimentally studied and theoretically investigated\cite{2223Co.wc,Wu2010,Zhu2010,Fuwa2010jacs, Fuwa2010ssc, Gunther2014, McCabe2014,2223Fe.Mayer,22221Fe,1221Fe.Hosono,2223Mn}.   A very good qualitative agreement  between theoretical calculations and experimental results has been reached.
 The Fe and Co  compounds are reported to possess Mott-insulating behavior and a small band gap of the Fe compound is found to be approximately $0.17-0.19$ eV\cite{Zhu2010, Lei2012,Wu2010}. Both Fe and Co are in the high spin state\cite{Gunther2014, McCabe2014, Fuwa2010jacs, Fuwa2010ssc, Wu2010}.  Experimentally, for the Fe compounds, the exact magnetic state   is still in debate\cite{Gunther2014, McCabe2014, Free2010} and  in the theoretical calculation\cite{Jin2016}, the G-type AFM state is obtained when U is taken to be 4.5eV and  another magnetic  state with propagation vector $\vec{Q} = (0.5, 0, 0.5)$ becomes competitive when U is less than 4.5eV. The Co-based compounds  are G-type AFM insulators in both experimental and theoretical studies\cite{Fuwa2010jacs}.


The G-type AFM state indicates the dominance of the nearest neighbor(NN) AFM exchange couplings, J$_1$, as shown in Fig.\ref{fig1}(c). The NN AFM interactions are mainly from direct exchange couplings between two d-orbitals.  As we will show later,  the situation becomes very different in the Ni-based compounds.  For the Ni-based ones,  the C-type of collinear antiferromagnetic striped state is favored, which indicates the dominance of the next NN (NNN) AFM exchange couplings. The NNN AFM interactions are from superexchange couplings through anion p-orbitals.  This magnetism trend in La$_2$O$_3$B$_2$Se$_2$ from Fe/Co to Ni is very similar to the one  in the family of materials related to iron-based superconductors. In the study of iron-based superconductors and related materials,  such as BaB$_2$As$_2$ (B=Cr, Mn, Fe),
the Cr and Mn-based compounds exhibit  the G-type AFM states but the Fe-based compounds exhibit the C-type AFM state. Thus, the trend from Fe/Co to Ni in La$_2$B$_2$Se$_2$O$_3$ exactly resembles the one from Cr/Mn to Fe in BaB$_2$As$_2$\cite{Zeng2017}.  As we have discussed in the beginning, the fact that only Fe-based compounds become high temperature superconductors upon doping or applying pressure has led us to argue  that only NNN superexchange couplings can lead to superconducting pairing\cite{ jpHusciencebult}.  Thus, here we have identified a new system to justify the assumption.

For the Ni-based compounds,  we calculate different chalcogen  compounds, La$_2$Ni$_2$M$_2$O$_3$  (M=S, Se, Te).  We list the optimized structural parameters in Table.\ref{structure2}.  The C-type Collinear AFM state is always favored in our calculations with different U values.  The ordered magnetic moments increase from 0.86$\mu_B $ at U=0 to 1.44$\mu_B$ at U=4eV.   When U is less than 2.0eV, the magnetic state is metallic. However, at U=3eV, it becomes an insulator with an insulating gap about 0.15eV and the gap increases further to about 0.5eV when U=4.0eV.

We follow the same procedure in ref.\cite{Zeng2017} to estimate the average effective magnetic interaction strength. By calculating  the energies in the different magnetic states, including  the ferromagnetic(FM) state, the G-type AFM state and the C-type stripe states, we can extract the average effective NNN AFM superexchange strengths for La$_2$B$_2$Se$_2$O$_3$(B = Fe, Co, Ni). The result is plotted in Fig.\ref{AFM}.  The NNN magnetic interactions in  the Fe/Co-based compounds are consistently  weak with the change of U. However, it is strongly antiferromagnetic in the Ni-based compounds under all calculations with different U values. The result proves that the superexchange AFM exchange interactions, namely the NNN AFM  interactions in  the Ni layer,  are  dominating.

\begin{table}[bt]
\caption{\label{structure2}%
Optimized structural parameters
of La$_2$Bi$_2$M$_2$O$_3$ by  GGA.}

\begin{ruledtabular}
\begin{tabular}{cccc}
       & La$_2$Ni$_2$S$_2$O$_3$& La$_2$Ni$_2$Se$_2$O$_3$ & La$_2$Ni$_2$Te$_2$O$_3$  \\
\colrule
 a(\AA) & 4.0197  &4.0834 & 4.1409  \\
 c(\AA) & 16.8855  &17.3271 & 18.7978  \\
 Ni-M(\AA)& 2.4920   & 2.5740 & 2.6920 \\
 M-M(\AA)& 2.9450    & 3.1360 & 3.4400    \\
 $\alpha$(Ni-M-Ni)& 107.543$^\circ$ & 104.959$^\circ$ & 100.559$^\circ$
\end{tabular}
\end{ruledtabular}
\end{table}

\begin{figure}[tb]
\centerline{\includegraphics[width=0.5\textwidth]{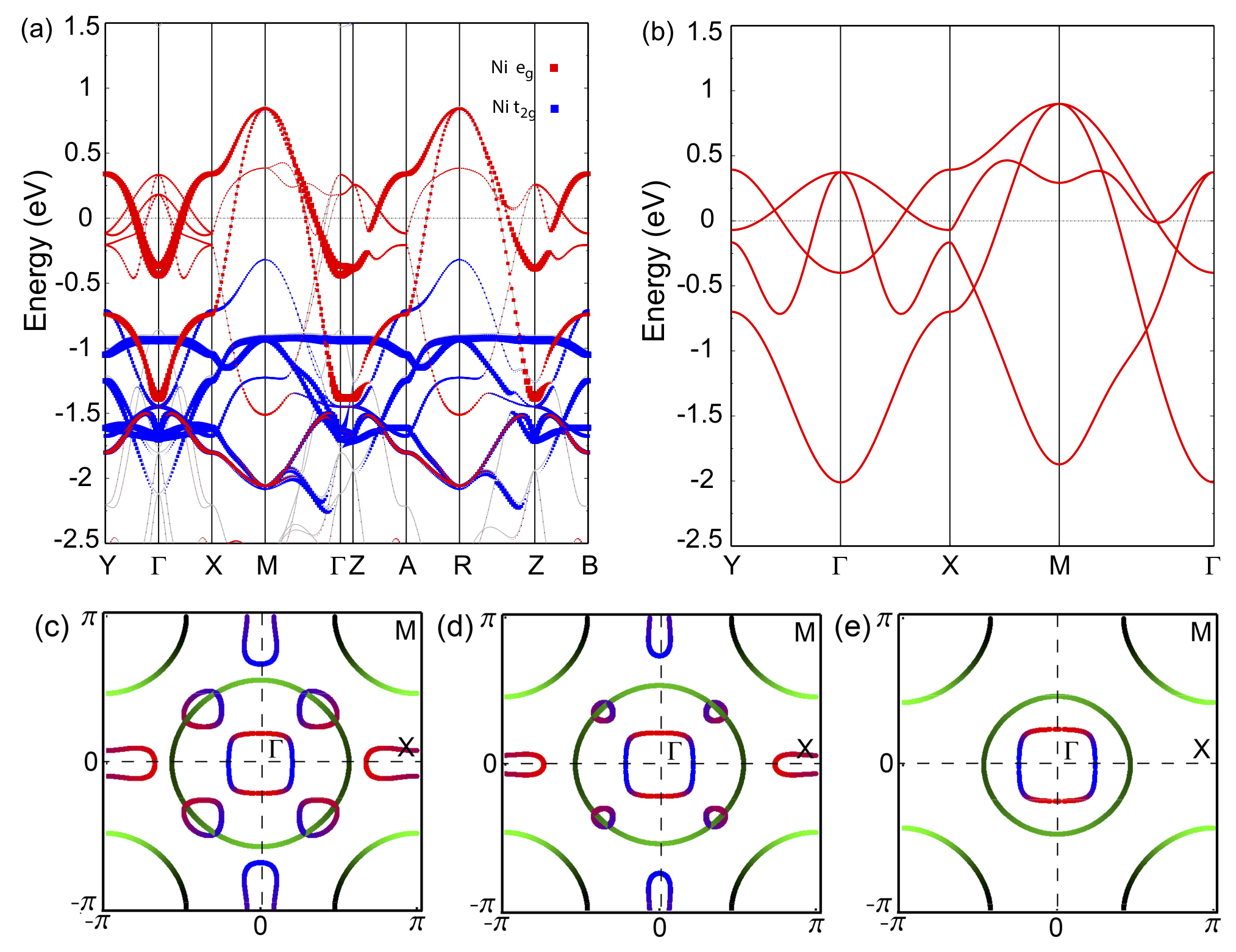}}
\caption{(color online) (a) The band structure of La$_2$O$_2$Ni$_2$Se$_2$O, The orbital characters of bands are represented by different grayscales; (b) The band structure of the effective model; (c), (d) and (e) show Fermi surfaces of the effective model at 0.2 electron doping, half filling and 0.28 hole doping, respectively. The orbital contributions  on Fermi surfaces are shown with different coded colors: Ni3 d$_{yz}$(red), Ni4 d$_{x^2-y^2}$ (green) Ni4 d$_{xz}$ (blue) and Ni3  d$_{x^2-y^2}$ orbitals (black).}
\label{allband}
\end{figure}

Second, we show that the presence of strong NNN AFM interactions is consistent with the presence of  the two near half-filling e$_g$ orbitals at Fermi energy  in the paramagnetic state.   We label the Ni atoms as shown in Fig.\ref{fig1}(b).   Specifically,  the e$_g$ orbitals of Ni(2,3) as indicated in Fig.\ref{fig1}(b) are d$_{x^2-y^2}$ and d$_{yz}$, and the e$_g$ orbitals of Ni(1,4) are d$_{x^2-y^2}$ and d$_{xz}$. The band structures  of La$_2$Ni$_2$M$_2$O$_3$  for different chalcogens are  very similar.  Therefore, in the following, we simply focus on La$_2$Ni$_2$Se$_2$O$_3$.  Fig.\ref{allband} shows the band structure   in which  different colors mark the orbital characters. The electronic structure is rather quasi-two dimensional. The main electronic physics is clearly  attributed to the monolayer Ni$_2$M$_2$O. The bands near the Fermi level are dominated by the e$_g$ orbitals. The $d_{x^2-y^2}$ orbital contributes to an electron pocket at the $\Gamma$ point and a hole pocket at the M point, and the $d_{xz,yz}$ orbital contributes a hole pocket at the $\Gamma$ point and an electron pocket at the X point. Both orbitals are near half-filling.

We can construct a  minimum effective tight binding model,   $H_0$,  to capture the two dimensional band structure near Fermi surfaces of the single layer Ni$_2$Se$_2$O.    We consider the base of the four e$_g$ orbitals at two different Ni sites (Ni3 d$_{x^2-y^2}$, Ni3  d$_{yz}$, Ni4 d$_{x^2-y^2}$, Ni4  d$_{xz}$). $H_0$ can be written as a $4\times 4$   matrix.  As the  band structures are decoupled between d$_{x^2-y^2}$ and  d$_{xz/yz}$ orbitals by symmetry,  the nonzero elements of $H_0$ matrix are given by
\begin{eqnarray}
\label{eq2}
& & H_{11}(k_x,k_y)=  \epsilon_{1}+2t^{11}_{xx}cos(k_x) +2t^{11}_{yy}cos(k_y)\nonumber
\\
& & H_{13}  = 4t^{13}_{xy}cos(0.5k_x)cos(0.5k_y)\nonumber
\\
& &H_{22}(k_x,k_y) = \epsilon_{2}+2t^{22}_{yy}cos(k_y)+2t^{22}_{yyyy}cos(2k_y)\nonumber
\\
& & +4t^{22}_{xxyy}cos(k_x)cos(k_y)+4t^{22}_{yyyyxx}cos(k_x)cos(2k_y)\nonumber
\\
& & H_{24}  =  -4t^{24}_{xy}sin(0.5k_y)sin(0.5k_x)
\end{eqnarray}
with $H_{3,3}(k_x,k_y)=H_{1,1}(k_y,k_x)$ and $H_{4,4}(k_x,k_y) = H_{2,2}(k_y,k_x)$.
 We use eV as the energy unit for all parameters. By fitting to the band structure of La$_2$O$_2$Ni$_2$Se$_2$O at the $k_z=0$ plane, we have $\epsilon_{1}=7.2218$  and $\epsilon_{2}=7.0804$  for the onset energy of $d_{x^2-y^2}$ and $d_{xz,yz}$. The corresponding hopping parameters in above equation are $t^{11}_{xx}=-0.3995$, $t^{11}_{yy}=-0.1264$, $t^{13}_{xy}=-0.2014$, $t^{22}_{yy}=0.1573$, $t^{24}_{xy}=-0.2705$, $t^{22}_{xxyy}=-0.0113$,$t^{22}_{yyyy}= 0.0656$, $t^{22}_{yyyyxx}= 0.0668$, where xy labels the hopping between two NN sites and xx(yy) labels the hopping between two next NN sites along x(y) directions.  The band structure of the effective $H_0$ is plotted in Fig.\ref{allband}(b) and  the typical Fermi surfaces at three different doping levels are also plotted in Fig.\ref{allband}(c)-(e). Both  well capture the DFT band structures of the e$_g$ orbitals. Summarizing above results on the magnetism and electronic structure, it is clear that  the Ni-based compounds meet our necessary condition for unconventional high T$_c$ superconductivity.

Finally, we provide a general analysis to qualitatively understand the possible superconducting states in the Ni-based compounds.  Instead of carrying out detailed theoretical calculations, we analyze the material based on  energy scale and general principle emerged in understanding both cuprates and iron-based superconductors.

First,   if there is an unified superconducting mechanism for unconventional high T$_c$ superconductors,  the maximum T$_c$  must represent the energy scale of the underlining model.   This argument is supported by experimental results in cuprates and iron-based superconductors. If we compare the maximum T$_c$ achieved in cuprates and iron-based superconductors, their ratio is about 3.  If we compare the effective hopping generated through anions, it is about 0.42eV in cuprates\cite{rmpPickett1989} and about 0.15eV in iron-based superconductors\cite{Kurohi2008}. Their ratio is also about 3.   Second, the pairing symmetries in cuprates and iron-based superconductors can be unified  within the Hu-Ding principle\cite{Hu_Ding, Davis29102013} which  states that in order to generating high T$_c$ superconductivity, the momentum space form factor of the superconducting pairing gap function which is determined by the AFM superexchange couplings must have large overlap with Fermi surfaces and the most favored pairing symmetry is the one which has the largest overlap strength\cite{Hu_Ding}.

 For the Ni-based compounds, we can find that the three NNN hoppings, $t^{11}_{xx}$, $t^{11}_{yy}$, and $t^{22}_{yy}$, are mediated through the p-orbitals of O/Se anions based on their signs.  Comparing their values with those in cuprates and iron-based superconductors, we can notice that the value of  $t_{xx}^{11}\sim 0.4eV$, which is mediated through oxygen,  is comparable to cuprates and the left two hopping parameters $t^{11}_{yy}$ and $t^{22}_{yy}$, which is mediated through Se, are comparable to those of iron-based superconductors. These three hoppings  are associated with three  AFM superexchange interactions, $J_2^{o,x^2-y^2}$, $J_2^{Se,x^2-y^2}$ and $J_2^{Se,xz/yz}$, specified in Fig.\ref{fig1}(c).

\begin{figure}[tb]
\centerline{\includegraphics[width=0.5\textwidth]{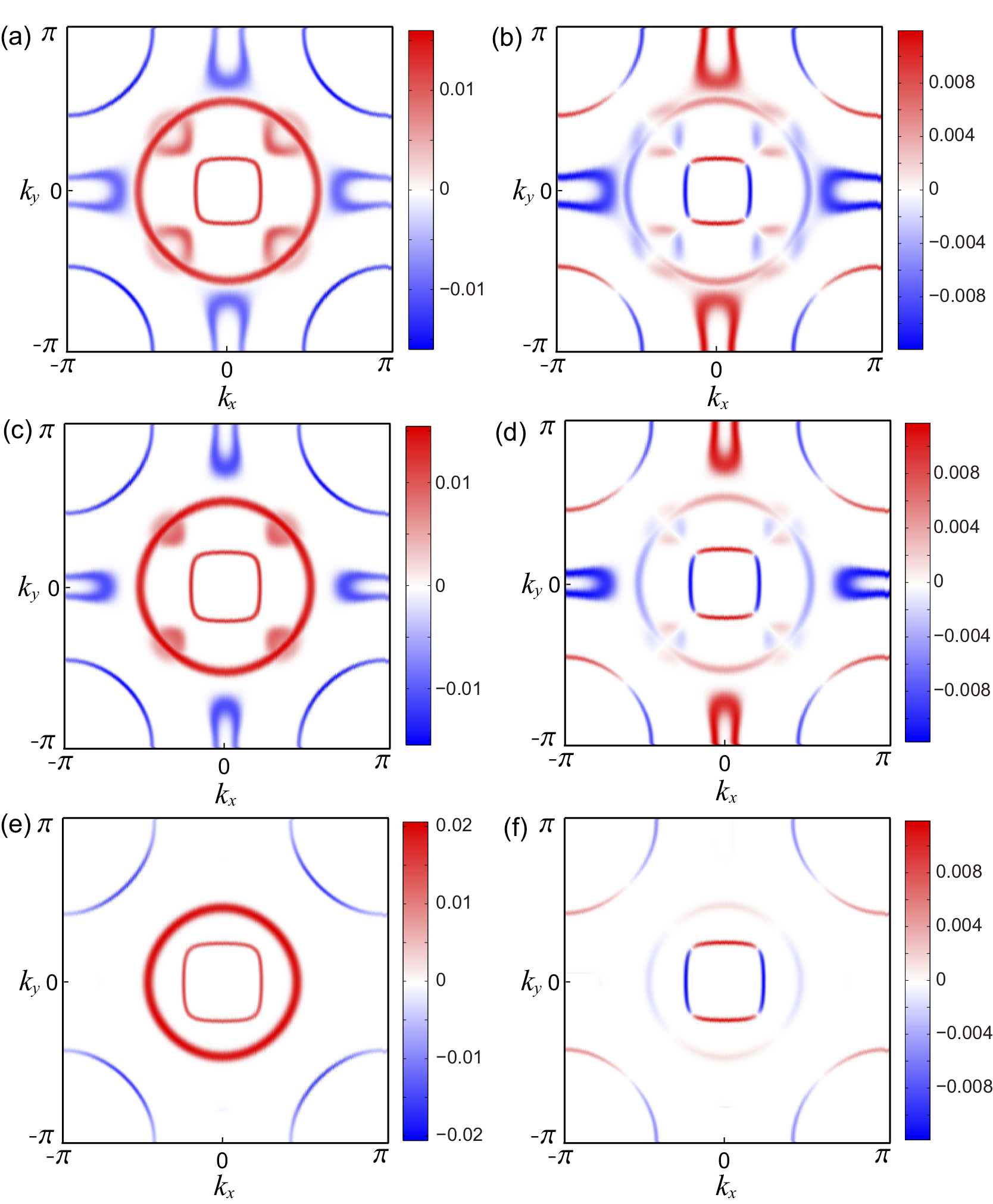}}
\caption{(color online) The superconducting gap structure for extended s-wave and d-wave based at the three different doping levels:(a, b), (c, d) and (e, f) are corresponding to 0.2 electron doping, half filling and 0.28 hole doping, respectively.}
\label{gapfunction}
\end{figure}

Based on the above energy scaling  and  assuming that the superconducting pairing is provided by short AFM exchange interactions,  we can use one superconducting gap parameter $\Delta_0$ to  write the pairing forms in momentum space  between two NN Ni(2,3) atoms being $ \Delta_0(cos(k_x)+ \frac{1}{3}cos(k_y))$  for $d_{x^2-y^2}$   and  $\frac{1}{3}\Delta_0cos(k_y)$ for $d_{yz}$, and between two NN Ni(1,4) atoms  being $\pm\Delta_0(cos(k_y)+\frac{1}{3}cos(k_x) )$  for $d_{x^2-y^2}$   and   $\pm \frac{1}{3}\Delta_0 cos(k_x)$ for  $d_{xz}$.
The positive and negative signs are corresponding to the extended s-wave and d-wave pairing symmetries.  With this choice of pairing functions, we can compare the gap values on different type of Fermi surfaces.  The gap values and the signs on the three type of Fermi surfaces in Fig.\ref{allband}(c) are plotted in Fig.\ref{gapfunction} by taking $\Delta_0=0.03eV$.  The overlap strength  between the form factors and the Fermi surfaces for the extended s-wave is 1.6, 2 and 5 times larger than the d-wave at 0.2 electron doping, half filling and 0.28 hole doping, respectively.  Therefore, the extended s-wave in hole doping region is reasonably much stronger than the d-wave. However, in the electron doping region, the d-wave is competitive to the extended s-wave, which suggests a possible rich physics diagram in this family of possible superconductors.  As the average energy scale   is higher than those of iron-based superconductors, the maximum T$_c$ should be higher than those of iron-based superconductors as well.

In summary, we have identified a possible new family of Ni-based high temperature superconductors, in which two e$_g$ orbitals  can be isolated near the d$^8$ filling configuration to carry  electronic physics. This key electronic character  has been missed in all known Ni-based compounds. Synthesizing this Ni-based family of compounds can provide us ultimate information to settle unconventional high T$_c$ mechanism.

It is  also worth   addressing a few points and mentioning  material perspectives.  First, in the search of Ni-based high temperature superconductors\cite{Anisimov1999},  the attention has been paid to  compounds with low valence Ni$^{+1}$ which resembles Cu$^{+2}$.
Ni$^{+1}$ is not a very natural valence configuration in chemistry  and can result in  valence orders. Moreover, as correlated electronic physics is required to be carried by d-orbitals, mixing with 4s-orbital in Ni$^{+1}$ can significantly weaken correlation effects. These reasons can be the major  fact why  unconventional high T$_c$ superconductors, so far, appear only in $+2$ valence transitional metal compounds.  Second, since the La$_2$B$_2$M$_2$O$_3$ structure is realized for B = Mn, Fe, and Co, it seems to be likely that the Ni-based analogue can also be synthesized.  However, further exploration on different anion combinations is needed to search for the best suitable conditions. For example, we may investigate new Ni-compounds by replacing chalcogens by pnictides or chloride and oxygen by fluorine. Third,  the Ni-compounds share many similarities on material and physical aspects to iron-based superconductors, including multi-orbital and  multi-Fermi surface pocket structures, we can investigate mechanisms or origins associated with other orders and degrees of freedom besides superconductivity, for example, the origin of nematicity.  Finally, we have not addressed effects on superconductivity from the interactions between two orbitals. This effect can result in much rich pairing pictures such as  broken time reversal symmetry superconducting states.  Moreover, although the maximum T$_c$ in this family should exceed those in iron-based superconductors from the energy scale argument, T$_c$ is expected to be sensitive to lattice parameters and bond angles as witnessed in iron-based superconductors so that external pressure can have a major effect on physical and superconducting properties.

{\it Acknowledgements}
The work is supported by the Ministry of Science and Technology of China 973 program (No. 2015CB921300, No.~2017YFA0303100), National Science Foundation of China (Grant No. NSFC-1190020, 11534014, 11334012),  the Strategic Priority Research Program of CAS (Grant No.XDB07000000), and the Key Research Program of the CAS(Grant No. XDPB08-1).


%
%
%

\end{document}